\documentstyle[prd,aps,epsfig]{revtex}

\author{S. Hegyi}

\title{H-function extension of the NBD: further applications}

\address{KFKI Research Institute for Particle and Nuclear Physics \\ of the
	Hungarian Academy of Sciences, \\ H-1525 Budapest 114, P.O. Box 49.
	Hungary}

\date{\today}

\begin{document}

\pagestyle{plain}
\pagenumbering{arabic}

\maketitle

\begin{abstract}
  The H-function extension of the Negative Binomial Distribution is
  investigated for scaling exponents $\mu<0$. Its analytic form is
  derived via a convolution property of the H-function. Applications
  are provided using multihadron and galaxy count data for $P_n$.
\end{abstract}


In a series of recent papers the present author developed a generalization
of the Negative Binomial Distribution~[1-3].
$P_n$ was obtained by means of the Poisson transform~[4]
\begin{equation}
	P_n=\int_0^\infty f(x)\,\frac{x^n}{n!}\,e^{-x}\,dx
\end{equation}
of the continuous probability density $f(x)$ which asymptotically yields the
KNO scaling form of $P_n$ (provided that $f(x)$ belongs to the family of
scale parameter generating density functions). Choosing $f(x)$ to be the
generalized gamma distribution~[5]
\begin{equation}
        f(x)=\frac{|\mu|}{\Gamma(k)}\,\lambda^k
        x^{\mu k-1}\exp\left(-\lambda x^\mu\right)
\end{equation}
with moments of the form
\begin{equation}
	\langle x^q\rangle=\int_0^\infty x^qf(x)\,dx=
	\left\{\begin{array}{ll}
	  \displaystyle
	  \frac{\Gamma(k+q/\mu)}{\Gamma(k)}\frac{1}{\lambda^{q/\mu}}
          	 & \quad\mbox{for $q/\mu>-k$} 	\\
	  \infty & \quad\mbox{otherwise}
	\end{array}\right.
\end{equation}
one obtains the above cited generalization of the NBD. In Eqs.~(2-3) we have
parameters $k>0$ (shape parameter), $\lambda>0$ (scale parameter), and
$\mu\neq0$ (scaling exponent). In the limit $\mu\to0$ and $k\to\infty$
the generalized gamma density converges to the log-normal law. The $k=1$,
$\mu>0$ special case of Eq.~(2) provides the Weibull distribution whereas
the $\mu=1$ case is, of course, the ordinary gamma distribution whose
Poisson transform yields the NBD.

In refs.~[1-3] the Poisson transform of the generalized gamma density
Eq.~(2) was investigated only for scaling exponents $\mu>0$. The analytic
form of $P_n$ was shown to be expressible in terms of Fox's H-function
(see refs.~[1,3] for a summary and ref.~[6] for details). Therefore we will
call this particular extension of the Negative Binomial Distribution
as HNBD for short. It takes the form
\begin{equation}
	P_n=\frac{1}{n!}\,\frac{\Gamma(k+n)}{\Gamma(k)}\;
	\theta^n\,(1+\theta)^{-k-n}\quad\mbox{for $\mu=1$}
\end{equation}
\begin{equation}
   P_n=
   \frac{1}{n!\,\Gamma(k)}\;{\sf H}^{1,1}_{1,1}
   \left[
      \,\theta\left|
      \begin{array}{c}
         (1-k,\; 1/\mu)             \\
         (n, 1)
      \end{array}
   \right]\right.
   \quad\mbox{for }\mu>1
\end{equation}
\begin{equation}
   P_n=
   \frac{1}{n!\,\Gamma(k)}\;{\sf H}^{1,1}_{1,1}
   \left[
      \,\frac{1}{\theta}\left|
      \begin{array}{c}
         (1-n,\; 1)             \\
         (k,\; 1/\mu)
      \end{array}
   \right]\right.
   \quad\mbox{for }0<\mu<1
\end{equation}
where ${\sf H}(\cdot)$ denotes the H-function of Fox and
$\theta=\lambda^{1/\mu}$ can be expressed in terms of the
average multiplicity $\langle n\rangle$ according to
$\theta=\langle n\rangle\Gamma(k)/\Gamma(k+1/\mu)$. The pure NBD
given by Eq.~(4) is the $\mu=1$ marginal case of Eq.~(5) for
$\langle n\rangle<k$ and of Eq.~(6) for $\langle n\rangle>k$~[3].
The $\mu\to0$, $k\to\infty$ Poisson transformed log-normal limit
of the HNBD lacks a representation in terms of known functions.

The main goal of the present Letter is to complete the derivation of the
Poisson transformed generalized gamma distribution by extending its
validity to negative values of the scaling exponent.
For $\mu<0$ we shall consider the Poisson transform of the probability
density
\begin{equation}
        f(x)=\frac{\mu}{\Gamma(k)}\,\lambda^k
        x^{-\mu k-1}\exp\left(-\lambda x^{-\mu}\right)
\end{equation}
with $\mu>0$. In the derivation of $P_n$ two particular cases of
${\sf H}(x)$ will be utilized,
\begin{equation}
	{\sf H}^{1,0}_{0,1}
   \left[
      \,ax\left|
      \begin{array}{c}
         -\!\!-             \\
         (0,1)
      \end{array}
   \right]\right.=e^{-ax}
\end{equation}
and
\begin{equation}
	\frac{1}{\Gamma(k)}
	{\sf H}^{1,0}_{0,1}
   	\left[
      	\,\frac{\theta}{x}\left|
      	\begin{array}{c}
         -\!\!-             \\
         (k,\,1/\mu)
      \end{array}
   \right]\right.=
	\frac{\mu}{\Gamma(k)}\,(\theta/x)^{\mu k}
        \exp\left(-[\theta/x]^{\mu}\right),
\end{equation}
the latter provides $xf(x)$ with $\theta=\lambda^{1/\mu}$ as before.
We shall also make use of the following two properties of
${\sf H}(x)$:
\begin{equation}
        x^r\,{\sf H^{m,n}_{p,\;q}}\left[\,x\left|
        \begin{array}{c}
          (a_1,\alpha_1),\ldots,(a_p,\alpha_p)      \\
	  (b_1,\beta_1) ,\ldots,(b_q,\beta_q)
        \end{array}
        \right]\right.=
        {\sf H^{m,n}_{p,\;q}}\left[\,x\left|
        \begin{array}{c}
   (a_1+r\alpha_1,\,\alpha_1),\ldots,(a_p+r\alpha_p,\,\alpha_p)     \\
   (b_1+r\beta_1,\,\beta_1) ,\ldots,(b_q+r\beta_q,\,\beta_q)
        \end{array}
        \right]\right.
\end{equation}
and
\begin{eqnarray}
	\int_0^\infty x^{-1}&&
	{\sf H^{m,n}_{p,\;q}}\left[\,ax\left|
        \begin{array}{c}
          (a_1,\alpha_1),\ldots,(a_p,\alpha_p)      \\
	  (b_1,\beta_1) ,\ldots,(b_q,\beta_q)
        \end{array}
        \right]\right.
	{\sf H^{M,N}_{P,\,Q}}\left[\,\frac{b}{x}\left|
        \begin{array}{c}
          (c_1,\gamma_1),\ldots,(c_P,\gamma_P)      \\
	  (d_1,\delta_1) ,\ldots,(d_Q,\delta_Q)
        \end{array}
        \right]\right.dx\nonumber\\=\;&&
	{\sf H^{m+M,n+N}_{\,p+P,\;q+Q}}\left[\,ab\left|
        \begin{array}{c}
          (e_1,E_1),\ldots,(e_{p+P},\,E_{p+P})      \\
	  (f_1,F_1),\ldots,(f_{q+Q},\,F_{q+Q})
        \end{array}
        \right]\right.
\end{eqnarray}
with parameters
\begin{eqnarray}
	\{(e_j,E_j)\}&&=(a_1,\alpha_1),\ldots,(a_n,\alpha_n),\,
		      (c_1,\gamma_1),\ldots,(c_P,\gamma_P),\,
		      (a_{n+1},\alpha_{n+1}),\ldots,(a_p,\alpha_p),
	\nonumber\\
	\{(f_j,F_j)\}&&=(b_1,\beta_1),\ldots,(b_m,\beta_m),\,
		      (d_1,\delta_1),\ldots,(d_Q,\delta_Q),\,
		      (b_{m+1},\beta_{m+1}),\ldots,(b_q,\beta_q).
\end{eqnarray}
For the conditions of validity of various properties of ${\sf H}(x)$,
see ref.~[6].
Absorbing $x^n$ into the lhs. of Eq.~(9) via identity~(10) one obtains
for $P_n$ defined by Eq.~(1) with $f(x)$ given by Eq.~(7)
the following integral:
\begin{equation}
	P_n=\frac{\theta^n}{n!\,\Gamma(k)}\int_0^\infty x^{-1}
	{\sf H}^{1,0}_{0,1}
   	\left[
      	\,\frac{\theta}{x}\left|
      	\begin{array}{c}
         -\!\!-             \\
         (k-n/\mu,\,1/\mu)
      \end{array}
   \right]\right.
{\sf H}^{1,0}_{0,1}
   \left[
      \,x\left|
      \begin{array}{c}
         -\!\!-             \\
         (0,1)
      \end{array}
   \right]\right.dx.
\end{equation}
Comparison with the convolution property Eq.~(11) yields
\begin{equation}
	P_n=\frac{\theta^n}{n!\,\Gamma(k)}\;{\sf H}^{2,0}_{0,2}
   \left[
      \,\theta\left|
      \begin{array}{c}
         -\!\!-              \\
         (0,1),\;(k-n/\mu,\; 1/\mu)
      \end{array}
   \right]\right.
\end{equation}
and using identity (10) we get
\begin{equation}
   P_n=
   \frac{1}{n!\,\Gamma(k)}\;{\sf H}^{2,0}_{0,2}
   \left[
      \,\theta\left|
      \begin{array}{c}
         -\!\!-              \\
         (n,1),\;(k,\, 1/\mu)
      \end{array}
   \right]\right.
\end{equation}
for the Poisson transform of the probability
density Eq.~(7). Above, $\theta$ is expressible in terms
of $\langle n\rangle$ according to
$\theta=\langle n\rangle\Gamma(k)/\Gamma(k-1/\mu)$ if $k>1/\mu$.
This can be seen from Eq.~(3) recalling that
$\langle x^q\rangle=\langle n(n-1)\ldots(n-q+1)\rangle$ for the Poisson
transform Eq.~(1). Accordingly, the factorial moments of Eq.~(15) may
diverge. The generating function
\begin{equation}
        {\cal G}(u)=\sum_{n=0}^\infty(1-u)^nP_n=
        \int_0^\infty e^{-ux}f(x)\,dx
\end{equation}
can also be derived via the convolution property Eq.~(11),
it turns out to be
\begin{equation}
   {\cal G}(u)=
   \frac{1}{\Gamma(k)}\;{\sf H}^{2,0}_{0,2}
   \left[
     \,u\theta\left|
      \begin{array}{c}
         -\!\!-      \\
         (0,1),\;(k,\; 1/\mu)
      \end{array}
\right]\right..
\end{equation}
Changing $u$ to $-it$ in Eq.~(17) the characteristic function
$\varphi(t)=\int_0^\infty e^{itx}f(x)\,dx$ is obtained for Eq.~(7).
The generating function of the HNBD corresponding to Eqs.~(5-6)
was determined in~[1].

With Eqs.~(15,17) we have completed the derivation of the Poisson transformed
generalized gamma distribution. According to a theorem of Bondesson, for
$|\mu|>1$ the characteristic function of Eq.~(2) is an entire analytic
function of finite order and therefore it must have complex zeroes which is
not permitted for infinitely divisible entire characteristic functions~[7].
Since the infinite divisibility of $f(x)$ is preserved by $P_n$ for Eq.~(1)
one can deduce that this important feature
holds for the HNBD if $0\leq|\mu|\leq1$.
In ref.~[1] it has already been demonstrated that the factorial cumulant
moments of the HNBD exhibit nontrivial (not alternating) sign-changing
oscillations for $\mu>1$ in accordance with the violation of infinite
divisibility of $P_n$. The effect is illustrated in Fig.~1
for shape parameter $k=1$ (Weibull case) which describes the inelastic
$pp$ and deep-inelatic $e^+p$ multipilicity data very well~[2,3].
Bondesson's theorem tells us that sign-changing oscillations of the
factorial cumulants may arise also for $\mu<-1$. But some preliminary
results indicate that this is not the case and the factorial cumulants
of the HNBD, when exist, remain positive for $\mu<-1$.

Analysing the experimental data for $P_n$ in different collision processes
it was found that the HNBD shape with $\mu<0$ is, although rare, not
completely absent. The best example is the full phase-space TASSO data in
$e^+e^-$ annihilations at $\sqrt s=34.8$~GeV~[8]. Most of the theoretical
models struggle in the description of this high statistics data set. One of
the few exceptions is ref.~[9] reporting
$\chi^2/\mbox{d.o.f.}=9.1/14$. It corresponds to the NBD fitted in two
matched domains of multiplicity $n$ with 3 fit parameters altogether.
The HNBD analysis was carried out with fixed shape parameter $k=1$ (details
of the fitting procedure can be found in~[1,3]). This particular case of the
HNBD produces $\chi^2/\mbox{d.o.f.}=7.9/15$ with parameters
$\langle n\rangle=13.595\pm0.050$ and $\mu=-10.053\pm0.388$.
The fit is illustrated in Fig.~2. At $\sqrt s=22$~GeV we obtained
$\chi^2/\mbox{d.o.f.}=4.9/11$, $\langle n\rangle=11.313\pm0.096$ and
$\mu=-12.745\pm1.449$. Despite of the relatively small statistics available
at lowest PETRA energy $\sqrt s=14$~GeV, the HNBD fit to the TASSO data fails
with $\chi^2/\mbox{d.o.f.}=20.4/10$. At top PETRA energy $\sqrt s=43.6$~GeV
the $\mu\to0$, $k\to\infty$ log-normal limit of the HNBD produces the best
quality fit ($\chi^2/\mbox{d.o.f.}=6.4/16$) similarly to previous results
obtained at the $Z^0$ peak~[1,3]. This change in the parametrization of the
best-fit HNBD indicates a slight narrowing of the KNO functions with
increasing~$s$ in the energy range investigated.
It is worth mentioning that the value of $\mu$ was found in the fitting
procedures to be extremely sensitive to tiny details of the experimental
$P_n$. Since the $e^+e^-$ multiplicity data exhibit approximate KNO
scaling and dominantly follow the $\mu=0$ Poisson transformed log-normal
shape, the $\mu<0$ type behaviour with divergent higher-order factorial
moments should not be taken too seriously.

Let us now turn our attention to a completely different application of the
HNBD. It is widely known that the $\mu=1$ special case, the pure NBD, yields
satisfactory results for galaxy count distributions. In ref.~[10] the
Zwicky catalogue of galaxy clusters was investigated. The probability of
finding $n$ galaxies in a Zwicky cluster ($\langle n\rangle\sim100$)
was found
to follow Eq.~(2) on a KNO plot with $\mu=1$ and $k\sim5$. We have
performed a similar analysis using the Lick galaxy catalogue which is one
of the most successful attempts to map the extragalactic sky. It is known
also as Shane-Wirtanen survey after the name of the authors of the
original catalogue~[11].
The galaxy counts are contained in $1246$ photographic plates
covering $\sim70$~\% of the sky, essentially the same region as the
Zwicky sample. Each plate covers $6^\circ\times6^\circ$
and consists of a $36\times36$ array of counts in
$10^\prime\times10^\prime$ cells. The catalogue contains
$\sim1.25\times10^6$ galaxies out of which cca. $34$~\% are double counts
because each plate overlaps its neighbour by at least $1^\circ$.
Most of the double counts can be eliminated by using only the center
$5^\circ\times5^\circ$ of each plate. In ref.~[12] several
multiplicative correction factors were determined to the raw counts of
Shane and Wirtanen. The gray-scale map of the corrected galaxy counts
presented in~[12] shows an impressive network of galaxies with lots of
clumps and filaments. An early computer model universe designed to
reproduce this fine structure can be found in~[13] where the frequency
distribution of galaxy counts in the $10^\prime\times10^\prime$ cells is
also displayed.

The count distribution of galaxies analysed here
was extracted from the Lick catalogue in ref.~[14].
The plate selection criteria, applied in statistical analyses
of the catalogue to exclude plates
near the galactic plane, resulted $467$ plates with $420300$ cells in total.
The average count in cells is $\langle n\rangle\sim1.4$. Despite of the
small mean of counts, cells with $n>20$ galaxies are not rare in the Lick
survey and the count distribution has a very long tail
extending well over $n/\langle n\rangle=10$. This is a sure
sign of highly non-Poissonian behaviour of fluctuations. In~[13] a
comparison of the Shane-Wirtanen map and a completely random distribution
of galaxies is shown --- the difference between
the visual appearance of the two patterns is huge.
Another way of demonstrating this strong departure is fitting the Lick
galaxy count distribution in the $10^\prime\times10^\prime$ cells by the
$\mu=1$, $k\to\infty$ Poisson limit of the HNBD. The fit is displayed in
the top left plot in Fig.~3. The best-fit Negative Binomial
with $k\sim2$ improves on the
Poissonian but the deviation remains substantial, see the top right plot
of the same figure. We mention that
the analysis of the Zwicky catalogue in~[10] also signals a similar
problem with the $\mu=1$
NBD/gamma model since the good agreement mentioned above
was obtained after truncating the high-$n$ tail of the data.
In ref.~[3] it was shown that long tailed distributions such as the $P_n$
in $p\bar p$ collisions at $\sqrt s=900$~GeV can be
successfully described by the
$\mu\to0$, $k\to\infty$ limit of the HNBD. Fitting the Poisson transformed
log-normal law to our galaxy data  one obtains further improvement, see
the bottom left plot in Fig.~3, but the agreement in the high-$n$ tail is,
again, unsatisfactory. The main difficulty of reproducing the Lick counts
lies in the fact that the distribution decays according to an inverse
power-law. The bottom right plot of Fig.~3 illustrates a HNBD fit
with negative scaling exponent $\mu$. As is seen this attempt
finally succeeds in describing the high-$n$ tail too. The best-fit
parameters are $\mu=-0.571\pm0.070$ and $k=8.512\pm0.447$.

Summarizing our results,
we have investigated the Poisson transform of the generalized
gamma distribution Eq.~(2) for negative values of the parameter $\mu$.
The analytic form of $P_n$ was derived by considering the Poisson
transform of the probability density Eq.~(7) with $\mu>0$. Making use of
the convolution property Eq.~(11) of the H-function we have obtained
Eqs.~(15) and (17) for the analytic form of $P_n$ and ${\cal G}(u)$.
Our result completes the specification of the Poisson transformed
generalized gamma distribution introduced in~[1] and developed in~[2,3].
Its application to multihadron and galaxy count data revealed
$\mu<0$ type behaviour in both cases. Nevertheless, it seems
to be unlikely that the HNBD shape with negative $\mu$
will be frequently encountered for the full phase-space multiplicity
distributions in particle and nuclear collisions. This feature
is more probable
for galaxy count data which often decay according to an inverse power-law
reflecting the scale-invariant distribution of galaxies in the universe.
Concerning multiparticle dynamics, similar behaviour of $P_n$ should be
searched in restricted domains of phase-space.

\section*{Acknowledgements}

I am indebted to A.S. Szalay and I. Szapudi for the discussions
concerning the Lick catalogue and for providing me their data.
T. Cs\"org\H o and G. Jancs\'o are acknowledged
for the many useful conversations on the subject.
This work was supported by the Hungarian Science Foundation under
Grant No. OTKA-T024094/1997.

\newpage

\mediumtext

\begin{figure}
\epsfig{file=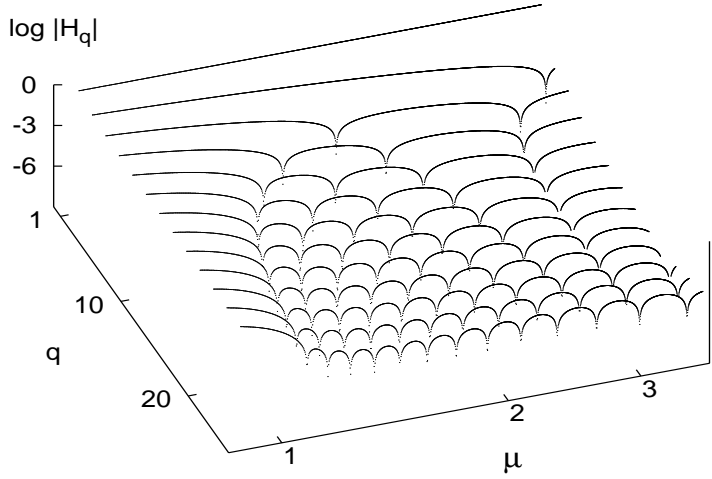,height=8cm,width=13cm}
\bigskip
\caption{Sign-changing oscillations of the factorial
cumulant-to-moment ratios $H_q$~[15] for the $k=1$ special case of HNBD.
Considering slices with  fixed rank $q$,
the neighbouring bumps along the $\mu$-axis
correspond to $H_q$ having opposite sign.
For $\mu\leq1$ the moment ratios are  always positive.
The $\mu$-scale is logarithmic and for clarity
$\log |H_q|$ is displayed only for odd ranks $q$.}
\end{figure}

\vskip1cm

\begin{figure}\hbox{\hskip.5cm
\epsfig{file=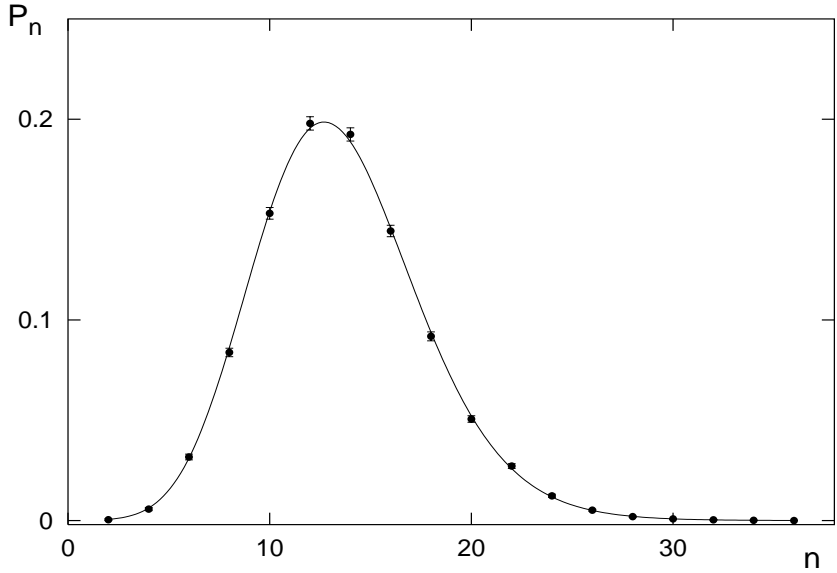,height=8cm,width=12cm}}
\bigskip
\caption{The best-fit HNBD to the full phase-space TASSO data at
$\sqrt s=34.8$~GeV~[8].
The fit parameters are: $k=1$ fixed, $\langle n\rangle=13.595\pm0.050$
and $\mu=-10.053\pm0.388$.}
\end{figure}

\newpage

\begin{figure}
\vskip-1cm\hbox{\hskip-.3cm
\epsfig{file=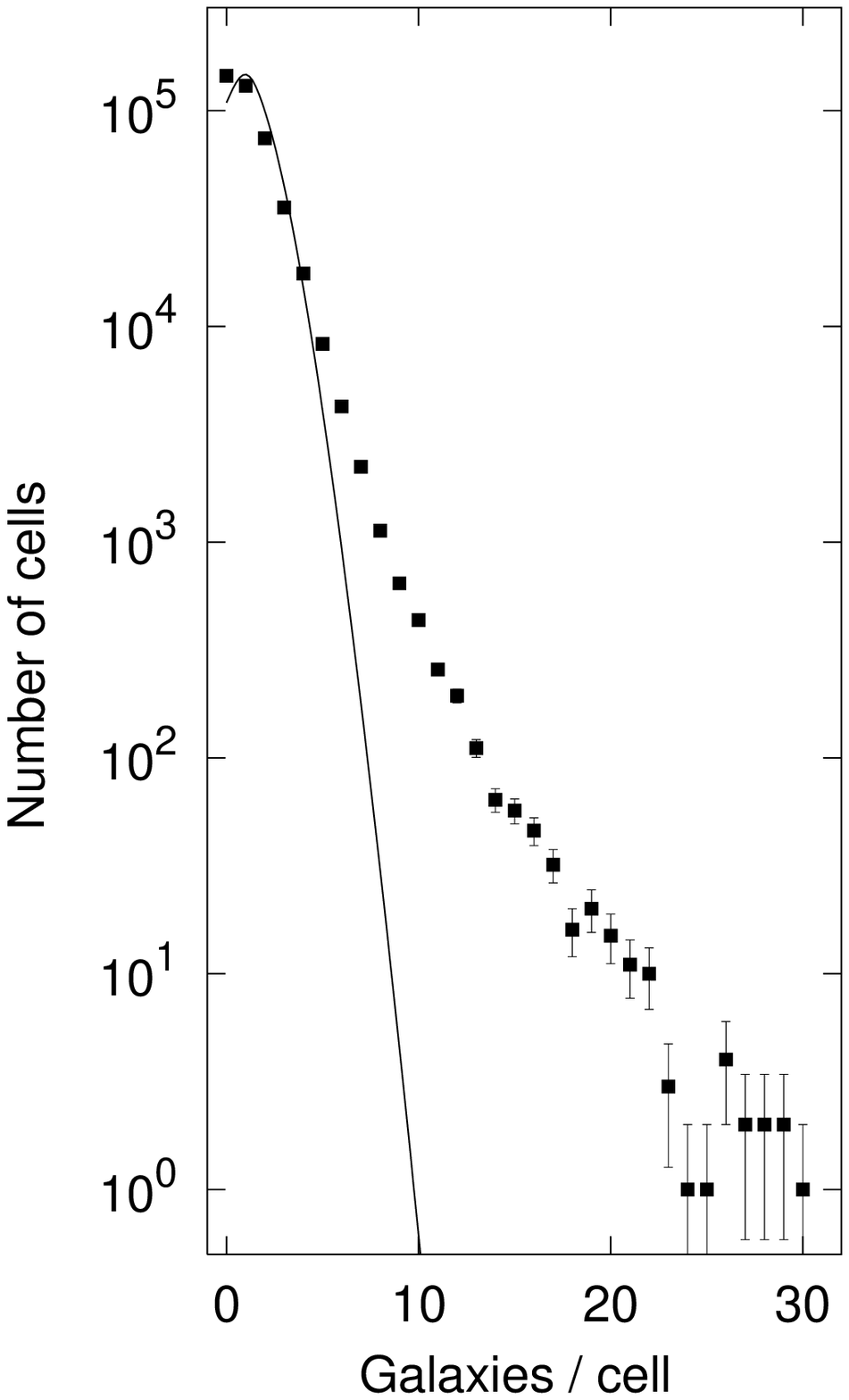,height=10.cm,width=7cm}\hskip-.5cm
\epsfig{file=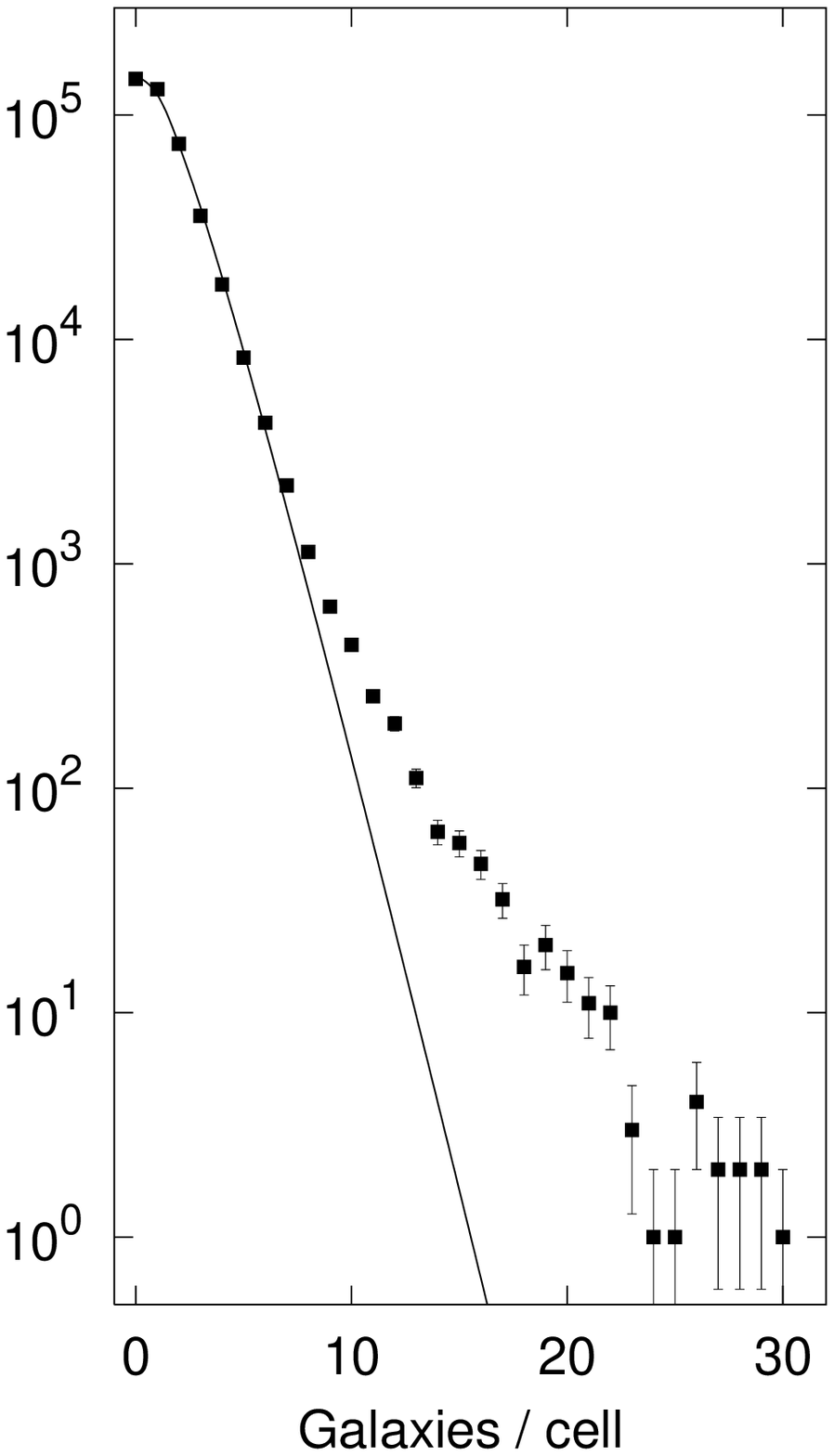,height=10.cm,width=7cm}}
\vskip.5cm\hbox{\hskip-.3cm
\epsfig{file=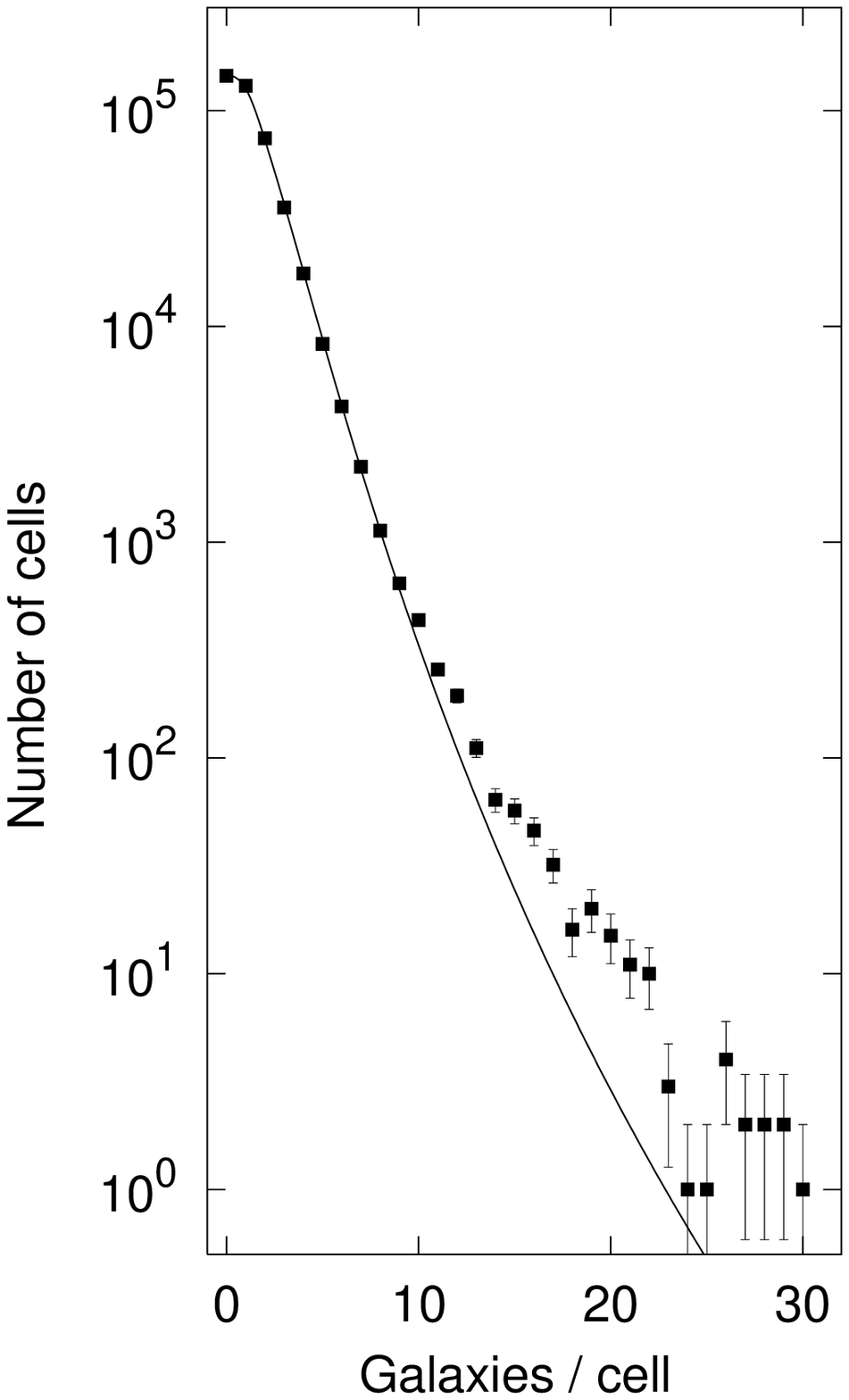,height=10.cm,width=7cm}\hskip-.5cm
\epsfig{file=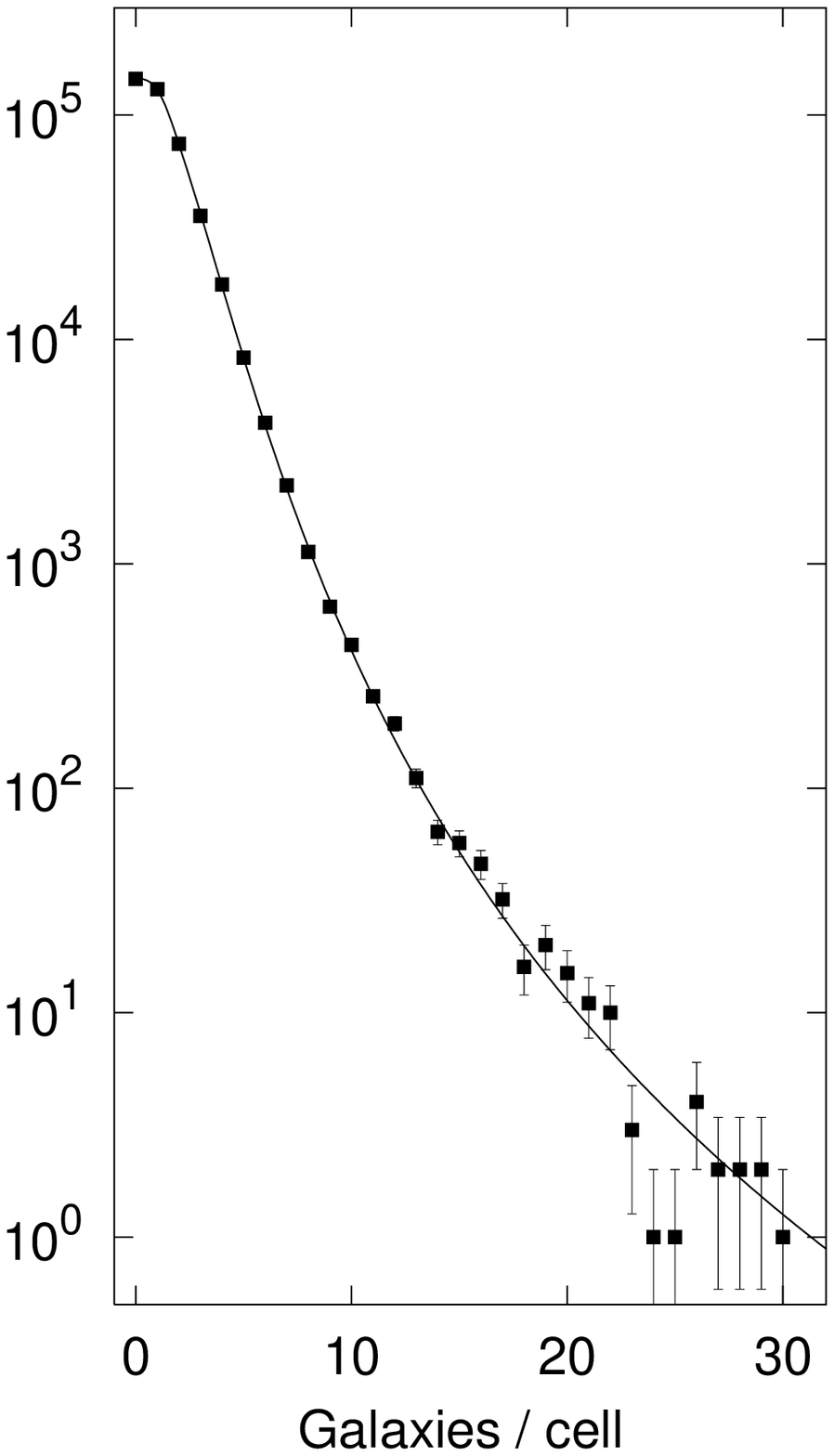,height=10.cm,width=7cm}}
\bigskip
\caption{Various special and limiting cases of HNBD fitted to the
Lick galaxy count distribution. Top left: best-fit Poisson ($\mu=1$,
$k\to\infty$); top right: best-fit Negative Binomial ($\mu=1$,
$k\sim2$); bottom left: best-fit Poisson transformed log-normal ($\mu\to0$,
$k\to\infty$); bottom right: best-fit~HNBD with parameters
$\mu=-0.571\pm0.070$ and $k=8.512\pm0.447$. The average count in cells is
$\langle n\rangle\sim1.4$. The errorbars represent Poisson errors.}
\end{figure}


\begin{references}
\bibitem{HNBD1} S. Hegyi, Phys. Lett. {\bf B387}, 642 (1996).
\bibitem{HNBD2} S. Hegyi, Phys. Lett. {\bf B388}, 837 (1996).
\bibitem{HNBD3} S. Hegyi, ``H-function extension of the NBD in the
light of experimental data'', hep-ph/9707322.
\bibitem{CS} P. Carruthers and C.C. Shih,
        Int. J. Mod. Phys. {\bf A2}, 1447 (1987).
\bibitem{JK} N.L. Johnson and S. Kotz,
        Distributions in Statistics Vol. 2.,
        Continuous Univariate Distributions (Wiley, 1970).
\bibitem{MS} A.M. Mathai and R.K. Saxena,
        The H-Function with Applications in
        Statistics and Other Disciplines \\ (Wiley Eastern, 1978).
\bibitem{BO} L. Bondesson, Scand. Actuar. J. 48 (1978).
\bibitem{TASSO} W. Braunschweig et al., Z. Phys. {\bf C45}, 193 (1989).
\bibitem{BH} S. Barshay and P. Heiliger, Z. Phys. {\bf C51}, 399 (1991).
\bibitem{CM} P. Carruthers and Minh Duong-Van,
			Phys. Lett. {\bf B131}, 116 (1983).
\bibitem{SW} C.D. Shane and C.A. Wirtanen,
			Publ. Lick Obs. {\bf XXII}, Pt. 1 (1967).
\bibitem{P1} M. Seldner, B. Siebers, E.J. Groth and P.J.E. Peebles,
			Astron. J. {\bf 82}, 249 (1977).
\bibitem{P2} R.M. Soneira and P.J.E. Peebles,
			Astron. J. {\bf 83}, 845 (1978); see also \\
	E.J. Groth, P.J.E. Peebles, M. Seldner and R.M. Soneira,
	Sci. Amer. {\bf 237}, 5 (1977).
\bibitem{SSP} I. Szapudi, A.S. Szalay and P. Bosch\'an,
			Astrophys. J. {\bf 390}, 350 (1992).
\bibitem{Dre} I.M. Dremin, Mod. Phys. Lett. {\bf A8}, 2747 (1993);
	Physics Uspekhi {\bf 37}, 715 (1994). \\
	I.M. Dremin and R.C. Hwa, Phys. Rev. {\bf D49}, 5805 (1994).
\end{references}
\end{document}